\documentclass[a4paper]{jpconf-kb}
\usepackage{amsmath}
\usepackage{iopams}
\usepackage{natbib}

\usepackage{multind}

\begin{document}

\title{Scaling properties of cosmic (super)string networks}

\author{C.J.A.P.\ Martins}
\address{Centro de Astrof\'{\i}sica, Universidade do Porto, Rua das Estrelas, 4150-762 Porto, Portugal}
\ead{Carlos.Martins@astro.up.pt}
\index{authors}{Martins, C.J.A.P.}

\begin{abstract}
I use a combination of state-of-the-art numerical simulations and analytic modelling to discuss the scaling properties of cosmic defect networks, including superstrings. Particular attention is given to the role of extra degrees of freedom in the evolution of these networks. Compared to the 'plain vanilla' case of Goto-Nambu strings, three such extensions play important but distinct roles in the network dynamics: the presence of charges/currents on the string worldsheet, the existence of junctions, and the possibility of a hierarchy of string tensions. I also comment on insights gained from studying simpler defect networks, including Goto-Nambu strings themselves, domain walls and semilocal strings.
\end{abstract}

\section{Introduction}

After a quest of several decades we now know, thanks to the recent LHC results \citep{Atlas,Cms} that fundamental scalar fields are among Nature’s building blocks. A pressing follow-up question is whether this Higgs-Kibble field has a cosmological role or, more generally, if there are further cosmological scalar field counterparts.

Cosmic defects \citep{Book,Encycl} are one of the possible signatures of this type of field. Quite apart from this specific motivation, the study of their formation and evolution is interesting since it can span an extremely wide range of energy, length and time scales, but despite this they are always described by the same underlying physical processes. When studying defect evolution in the early universe one is looking at microscopically small defects, which can nevertheless play an important role on cosmological scales. Examples of other contexts where defects have been studied (and seen) include liquid crystals \citep{Chuang} and also tafoni \citep{Tafoni} which may be seen on the seaside and have also been identified on Mars.

This article provides a brief introduction to (and recent results from) analytic and numerical methods used to study the evolution and cosmological consequences of these networks. (Other contributions in these proceedings describe related aspects.) In particular, in a cosmological context one's main interest is often to look for (late-time, asymptotic) scaling solutions, so this will be our main emphasis in what follows.

Throughout this contribution, when saying that a given defect network is (or is not) scaling, we will be referring to the following 'linear' scaling solutions, for a characteristic lengthscale $L$ and a root-mean squared velocity $v$
\begin{subequations}\label{scaling0}
\begin{align}
L & \propto t \,,\label{scaling0a}\\
v & \propto {\rm const}\,.\label{scaling0b}
\end{align}
\end{subequations}
The physical relevance of linear scaling (for cosmic strings and superstrings, as well as for semilocal strings and global monopoles) is that it corresponds to the case where the string density remains a constant fraction of the background density, and hence can in principle paly a number of useful cosmological roles. If the characteristic lengthscale grows faster than this the network will eventually disappear, while if it grows more slowly it will eventually dominate the universe.

It should also be emphasised that we will be discussing {\it cosmic} defect networks: the universe is assumed to be expanding, and often one assumes for simplicity that the scale factor grows as a generic (but constant) power of physical time
\begin{equation}
a(t)\propto t^\lambda\,.
\end{equation}
This is important especially when one is doing numerical simulations of defect networks. It is well known that some of the solutions that we shall describe only exist in expanding universes. If one simulates networks in Minkowski spacetime, different results will be found. (An example of this will be discussed below.)

Having said that, we should also emphasise that the methods we shall describe are in principle applicable to any defect network. Indeed they can be (and have been) applied, {\it mutatis mutandis}, to defect networks in other contexts \citep{Condmat1,Condmat2}.

\section{Models for defect networks}

Broadly speaking, the {\it ethos} of analytic modelling of cosmic defect networks is to start from statistical physics but turn it into thermodynamics. This approach was introduced by \cite{Kibble}, and formulated in a fully quantitative way by \cite{VOS1,VOS2,VOS3}. In other words, one starts out from the knowledge of the microphysics of the defects (given, for the case of strings, by the well-known Goto-Nambu action) an uses this to obtain evolution equations for macroscopic quantities, (possibly) suitably averaged over the network.

The two main macroscopic quantities of interest are a characteristic lengthscale and an averaged (root-mean-square) velocity. The physical interpretation of the latter is fairly straightforward. As for the first, one could in principle interpret it in several different ways:
\begin{itemize}
\item The inter-string separation (in other words, a measure of the string density);
\item The correlation length (in the usual physics sense) of the strings;
\item The curvature radius of the strings.
\end{itemize}
In the context of an analytic one-scale model, one assumes that these are all equal (or at least of the same order). In the case of cosmic strings, numerical simulations have shown that this is a good assumption for scaling networks. Having said that, one can certainly imagine (and simulate) networks in which the three are significantly different (or evolve differently).

These two macroscopic quantities are the bare minimum required in order to have a quantitative defect evolution model. Naturally, models with additional degrees of freedom (such as charges and/or currents on the string worldsheet) can be constructed. The model complexity will of course increase, but the same formalism will still apply.

Naturally, having an analytic model for the evolution of defect networks has considerable computational advantages, Nevertheless, one must bear in mind that these come at a price. In going from the microphysics to the macrophysics one is necessarily forced to introduce phenomenological parameters, and these cannot be calculated {\it ab initio}. The only way to determine their values is therefore to measure them directly from numerical simulations, thus calibrating the model. In other words, a detailed (and quantitative) understanding of the evolution of cosmic defects necessarily requires a combination of analytic modelling and high-resolution numerical simulation.

A detailed and rigorous derivation of the evolution equations of the analytic model for cosmic strings can be found in \cite{VOS1,VOS2,VOS3}; a simplified alternative derivation (which has the advantage of being applicable to defect networks of any dimensionality) can be found in \cite{Monopoles}. For defects with an $n$-dimensional worldsheet (monopoles correspond to $n=1$, strings to $n=2$, and so forth) the energy density $\rho$, symmetry breaking scale $\eta$ and characteristic lengthscale $L$ are related via
\begin{equation}
L^{4-n}=\frac{\eta^n}{\rho}\,.
\end{equation}
and the evolution equations for $L$ and the velocity $v$ are
\begin{subequations}\label{vos0}
\begin{align}
(4-n)\frac{dL}{dt} & = (4-n)HL +v^2\frac{L}{s}+\delta\epsilon\,,\label{vos0a}\\
\frac{dv}{dt} & = (1-v^2) \left(f-\frac{v}{s}\right)\,.\label{vos0b}
\end{align}
\end{subequations}
Here $H={\dot a}/a=\lambda/t$ is the Hubble parameter, $f$ parametrises driving forces (which may be due to curvature, to long-range forces, etc), $\delta\epsilon$  describes energy losses due to interactions, and $s$ is a damping lengthscale (including Hubble damping, friction due to particle scattering, and possibly other mechanisms). Naturally all the art in this modelling depends on how well one can describe these various terms.

It is interesting to point out that cosmic defect network evolution is to a large extent about energy flow from large to small scales (but other mechanisms, such as topological constraints, can also play a role). Defects gradually enter the horizon as the universe expands, and this energy is then lost (through loop production, annihilation or other mechanisms) on substantially small scales. In this sense, defect evolution is similar to a Richardson cascade, which is also discussed (in other contexts) elsewhere in these proceedings.

\section{A walk in the zoo}

This analytic model can be used to quantitatively confirm more qualitative work that had been done in the past, as well as to study defect evolution in contexts that were beyond the reach of those simpler analyses. In this section we briefly go through one example, by briskly but thoroughly discussing the evolution of networks containing monopoles (on their own or  attached to various numbers of strings).

The evolution of networks of local monopoles has been shown to be very sensitive to its available annihilation mechanisms \citep{Preskill,Monopoles}. On the other hand, scaling is quite generic for the case of global monopoles, in good agreement with the first generation of numerical simulations \citep{Bennett,Yamaguchi}. It would be highly desirable to have larger (and better resolution) numerical simulations to obtain a better calibration of the model.

Next, consider the case of networks of monopoles attached to one string. These are often referred to as hybrid networks, and they emerge as a result of the symmetry breaking sequence
\begin{equation}
G\rightarrow K\times U(1)\rightarrow K\,;
\end{equation}
the monopoles are formed at the first phase transition, while at the second one the strings form connecting monopole-antimonopole pairs. In most circumstances these networks annihilate very quickly (in much less than one Hubble time after string formation) \citep{Hybrid}. Nevertheless, with considerable fine-tuning one can make them survive much longer than a Hubble time: one requires that both phase transitions occur around the Planck scale, and that the network is not subject to friction due to particle scattering.

Finally we have the case of networks of monopoles attached to two or more strings. There result from the symmetry breaking sequence
\begin{equation}
G\rightarrow K\times U(1)\rightarrow K\times Z_n\,;
\end{equation}
in the former case ($n=2$) they are known as necklaces, while for $n\ge3$ they are known as lattices. In both cases it can be shown that they usually reach linear scaling \citep{Lattices}. However, other scaling solutions can also exist, depending on the universe's expansion rate and the network's energy loss mechanism(s).

\section{Cosmic strings and walls}

Attractor (large-scale) linear scaling solutions for cosmic strings are well established, and in particular they have been confirmed by everyone that simulates them---see \citet{Book} and \citet{Encycl} for overviews. Nevertheless one should bear in mind that large-scale scaling and scaling of the network's small-scale features are two conceptually distinct processes, and one can occur without the other. A further issue (which is not yet fully resolved) is the behaviour of the loop distribution; this is not crucial for our present purposes (although it is discussed elsewhere in these proceedings), but it is highly relevant for gravitational wave bounds.

In the case of cosmic strings the analytic model has been calibrated using both field theory \citep{Moore,Condmat2} and Goto-Nambu simulations \citep{Condmat2,Fractal}. Although this has been highly successful (and led to the adoption of the model for making predictions in a number of contexts, including a part of the Planck data analysis pipeline), it should be kept in mind that the simulations also indicate that extra degrees of freedom (in the form of small-scale 'wiggles') are clearly important if one requires more detailed modelling. It is equally clear that these will be crucial for the case of cosmic superstrings.

We will briefly illustrate the above point with two examples drawn from \cite{Fractal}. First, one often uses the intuitive picture that strings are Brownian on scales of the correlation length, but this is quantitatively not correct (for any sensible definition of the string correlation length). Strings are only Brownian outside the horizon; at scale of correlation length they can more accurately be described as a self-avoiding random walk. This is a consequence of the string intercommutings (and the ensuing loop production).

Second, it is clear from Goto-Nambu simulations in expanding universes that string velocities are anticorrelated on the scale of the correlation length. This anticorrelation depends on the expansion rate (it becomes stronger as the expansion rate is increased), and---crucially---it doesn't exist in Minkowski space. Physically this is easy to understand: it is due to string intercommutings, and specifically it is a manifestation of the fact that loop production is radically different with and without expansion. (The latter can also be confirmed by looking at loop family trees from simulations with and without expansion). Anticorrelations are easy to see in Goto-Nambu simulations but can't be seen in field theory simulations, which don't have enough spatial resolution. Additionally, accurate velocity measurements are notoriously difficult in field theory simulations. These anticorrelations are important for a quantitative assessment of the astrophysical signatures of these networks. This is likely to be the main factor explaining the difference (of more than a factor of two) between bounds on the string tension obtained from field theory and Goto-Nambu simulations.

One should also bear in mind that numerical simulations often assume a constant expansion rate (which we have denoted $\lambda$), but the real universe goes through several different epochs with different expansion rates. A string network will approach a scaling solution if it evolves sufficiently in the radiation or matter eras (or indeed any epoch with a constant $\lambda$). However, the universe goes through a relatively slow radiation-to-matter transition, and for GUT-scale strings the matter-era attractor solution would be approximately reached at a redshift just below $z\sim1$, which happens to coincide with the onset of dark energy domination (in which context strings no longer scale). The end result is that throughout the entire period of the universe that is relevant for structure formation and CMB anisotropies a cosmic string network is {\it never} in a linear scaling solution, although it is always relatively close to it. This is important if one wants to make quantitative predictions for their observational consequences---Planck is again a good example.

In passing we can also discuss the case of semilocal strings (which are further discussed in Ana Ach\'ucarro's contribution). These are not purely topological, and can end in a cloud of energy which one can effectively think of as a global monopole.  They are stable, both perturbatively and to semi-classical tunnelling into vacuum.

The study of these networks is significantly more complex than that of standard cosmic strings, mostly due to the fact that the dynamics of semilocal segments is highly non-trivial, and further work on this topic is currently in progress. In has already been shown that a scaling attractor exists for some range of model parameters \citep{Nunes}. However there can also be regimes where the whole network can disappear, and those where only the monopoles disappear (leaving behind a standard Goto-Nambu string network). Preliminary comparisons with numerical simulations are highly encouraging (a more detailed analysis will be available soon), but the main bottleneck is again measuring the defect velocities.

Also as brief passing remark, we note that there is a possible alternative analytic description of a semilocal network: instead of a network of local strings attached to global monopoles, one can instead consider a network containing two types of strings: the standard ones and also 'gradient strings' accounting for the fact that depending on the field configurations the segments might want to grow and join other segments (instead of shrinking and annihilating). This is, to some extent, analogous to the case of superstring networks.

Finally, for the case of the simplest domain wall networks (described by a single scalar field), an attractor large-scale linear scaling solution is also now well established \citep{Walls1,Walls2}, despite earlier claims of deviations from scaling which are now understood to be due to limited dynamical range of earlier simulations. These simulations have also been used to obtain an accurate calibration of the analytic model for domain walls: in this case the two free model parameters have been found to have the values
\begin{equation}
c_w = 0.34\pm0.16\,
\end{equation}
for the network energy losses term and
\begin{equation}
k_w = 0.98\pm0.07\,
\end{equation}
for the walls curvature parameter. Although the cosmological presence of domain walls is very tightly constrained (since they tend to have pathological consequences), domain walls provide a simple and extremely useful toy model. Being fairly simple to simulate, they allow one to study in detail the various physical processes contributing to defect evolution.

\section{How robust is scaling?}

So far we have been considering the simplest defect models, and in most cases finding that a scaling solution exists. One may ask whether this result still holds for more complex models. As we will see, the answer seems to be yes: the scaling solution is indeed quite robust and it persists in many cases, although there are certainly mechanisms that can suppress it.

We start by pointing out that explicit energy losses are not absolutely necessary for scaling in a cosmological context: the expansion itself may provide sufficient damping. For strings, scaling will occur for any expansion rate $\lambda>1/2$, regardless of energy losses \citep{Nonint} for domain walls, the analogous threshold is $\lambda >1/4$. Among other things this implies that for strings to scale in the matter era loop production not needed for scaling, as is known since the first generation of numerical simulations in the early 1990s. That said, note that the above paper also shows that in some string-dominated universes the correlation length grows as $L\propto a\propto t$, so this is formally a linear scaling solution, although its physical properties are somewhat different from the standard ones.

The next question has to do with the role of additional degrees of freedom on the string worldsheet. Perhaps the simplest such example is that of a charge, and it has recently been shown by \cite{Oliveira} that this need not stop scaling. More specifically these solutions fall into two regimes. If the expansion rate is large enough, the charge gets diluted and the standard scaling is recovered. Conversely if the expansion rate is small, the charge stays on the strings and there's no linear scaling solution: instead the correlation length grows more slowly and velocities decay. The transition between these two regimes depends on the balance between network energy losses and the expansion rate: with no energy losses, transition occurs for the matter era ($\lambda=2/3$), but if energy losses are present this threshold will be lower, and for large enough losses it will be below radiation, meaning that the presence of a charge on the string need not have dramatic consequences.

Similarly, the presence of string junctions need not prevent scaling. This result has in fact been known (even if only in a specific class of models) for a long time \citep{Tanmay}. It has also subsequently been obtained for other classes of models \citep{Nfields}, and it can occur for any number of coupled scalar fields (under the assumption that all walls have the same tension). A more general exploration of the role of defect junctions is in progress, and will be discussed elsewhere.

Finally, even the assumption of equal tensions can be relaxed: a hierarchy of different tensions need not prevent scaling. Again this result has been known for more than a decade, at least for one specific class of models \citep{Mcgraw}. The conditions under which it can be generalised are also under current study, and will be discussed elsewhere.

\section{Realistic cosmic superstrings}

From the discussion above, it should be clear that our approach to modelling realistic cosmic superstring networks is to isolate the various different physical processes that contribute to the dynamics of network, and try to understand the role of each one of them by using suitably simplified (toy) models. One can then put together this information into more accurate models and see what these imply for superstrings.

A number of these physical processes are under control, in the sense that their role is well understood. Specifically, three key examples are
\begin{itemize}
\item The dynamical role of monopoles and junctions, as briefly discussed above.
\item The role of extra dimensions (if applicable) and topology, as first discussed in \cite{Tasos1}.
\item The role of the intercommuting probabilities, first considered in a simple way in \cite{Nonint} and then quantified using numerical simulations in \cite{Tasos2}.
\end{itemize}

On the other hand, another set of relevant physical process have been identified but their understanding still requires more work (which is being done,  by many people). Again I will highlight three examples
\begin{itemize}
\item The role of additional degrees of freedom (such as charges and currents) needs to be understood in more detail. In these models a second lengthscale must necessarily be introduced, so these will no longer be 'one-scale' models. There are in principle several different ways to do this, which may be relevant for different contexts.
\item The behaviour of networks with various tension hierarchies needs to be better understood. A naive approach that has been followed is simply to treat the network as a sum of sub-networks (one for each tension) and to add together 'one-scale' models for each of these, in a minimally coupled way. However this approach (even if it is done in a way that conserves energy, which is not always the case) has its limitations, and may well miss some of the relevant physics.
\item More related to the networks' observational consequences, there is the role of the non-trivial velocity correlations. As mentioned above these exist for Goto-Nambu strings, and are expected to persist (with some differences) in superstring networks. Incidentally, there has been a fair amount of recent work on the so-called kinematic constraints on cosmic superstrings (some of which is discussed elsewhere in these proceedings); although this is certainly a possible approach, to my mind it it not ideal: physically the observable effect of the presence of the junctions should be to introduce non-trivial velocity (anti)correlations on the network, and the natural way to model is to use fractal-like tools.
\end{itemize}

Although a fully self-consistent analysis of the scaling properties of cosmic string networks still remains to be carried out, partial results from various groups (and our own ongoing work) do suggest that full scaling of all the components of the network can certainly occur in significant regions of parameter space. However, there are also regions of parameter space where only the lowest tension string scales (and the heaviest ones decay), and others where no component scales and the network can eventually (at least in some cases) dominate the universe's energy budget. Further work on these issues is ongoing. Which regions of parameter space are natural (or indeed realistic) is a different issue, which is beyond the scope of this analysis.

\section{Conclusions}

Analytic and numerical work in the past three decades has gradually established the result that scaling attractor solutions are ubiquitous for cosmic defect networks. At least at the qualitative level, the physical reasons behind this are clear: they stem from the usual energy minimisation mechanisms. Having said that, the solution is certainly not universal: in some case thee are additional physical mechanisms that naturally act to suppress it, while in others it can be suppressed at the cost of fine-tuning or otherwise violent assumptions.

An accurate description of the evolution of cosmic defect networks requires quantitative analytic models calibrated by high-resolution and physically realistic numerical simulations. This goal has been achieved for the simplest defect networks (in particular for domain walls and cosmic strings) and will be achieved soon for others (including monopoles and semilocal strings). In the longer term, the goal will be to apply these methods to cosmic superstring networks: currently only toy models of these network have been studied (both analytically and numerically). and more realistic approaches are highly desirable.

Naturally, the ultimate goal of all this work is to improve astrophysical and cosmological constraints. Current astrophysical bounds largely rely on toy-model assumptions, and (as shown by current discrepancy between bounds obtained from Goto-Nambu and field theory simulations) are not more accurate than a factor of two. Improvements are certainly needed, so that more robust analyses can be done as more Planck data becomes available.

\ack
Many interesting discussions with my collaborators in the work discussed herein (Ana Ach\'ucarro, Tasos Avgoustidis, A. Mafalda Leite, Andr\'e Nunes, Miguel Oliveira, Paul Shellard and Jon Urrestilla) as well as with the rest of CAUP’s Dark Side have shaped my views on this subject, and are gratefully acknowledged.

Our numerical simulations were performed on the COSMOS Consortium supercomputer (within the DiRAC Facility, jointly funded by STFC and the Large Facilities Capital Fund of BIS-UK) and the Advanced Computing Laboratory at University of Coimbra. 

This work was done in the context of the project PTDC/FIS/111725/2009 from FCT. CJM is also supported by an FCT Research Professorship, contract reference IF/00064/2012, funded by FCT/MCTES (Portugal) and POPH/FSE (EC).

\bibliographystyle{jfm2}
\bibliography{ini}

\end{document}